\documentclass[twocolumn,showpacs,aps,prl,superscriptaddress]{revtex4-1}
\usepackage{graphicx}
\usepackage{dcolumn}
\usepackage{amsmath}
\usepackage{epsfig}
\usepackage{subfloat}
\usepackage{colordvi}
\usepackage{color}
\usepackage{hhline}
\usepackage{multirow}
\usepackage{subfigure}
\usepackage{rotating}
\usepackage{slashbox}
\usepackage{hyperref}
\usepackage{url}

\RequirePackage{xspace}

\def\pep2{PEP-II}

\usepackage{relsize}
\def\babar{\mbox{\slshape B\kern-0.1em{\smaller A}\kern-0.1em
    B\kern-0.1em{\smaller A\kern-0.2em R}}}

\newcommand{\gev}{\ensuremath{\mathrm{\,Ge\kern -0.1em V}}\xspace}
\newcommand{\mev}{\ensuremath{\mathrm{\,Me\kern -0.1em V}}\xspace}
\newcommand{\gevc}{\ensuremath{\mathrm{\,Ge\kern -0.1em V}/c}\xspace}
\newcommand{\mevc}{\ensuremath{{\mathrm{\,Me\kern -0.1em V}}}\xspace}
\newcommand{\gevcc}{\ensuremath{{\mathrm{\,Ge\kern -0.1em V}}}\xspace}
\newcommand{\gevccs}{\ensuremath{{\mathrm{\,Ge\kern -0.1em V^2\!/}c^4}}\xspace}
\newcommand{\mevcc}{\ensuremath{{\mathrm{\,Me\kern -0.1em V}}}\xspace}

\def\kk         {\ensuremath{KK}\xspace} 
\def\kpi         {\ensuremath{K\pi}\xspace} 
\def\pipi         {\ensuremath{\pi\pi}\xspace} 
\def\akk         {\ensuremath{A_{KK}}\xspace}   
\def\akpi         {\ensuremath{A_{K\pi}}\xspace}
\def\apipi        {\ensuremath{A_{\pi\pi}}\xspace}

\def\epem       {\ensuremath{e^+e^-}\xspace}

\def\tautau     {\ensuremath{\tau^+\tau^-}\xspace}

\def\qbar  {\ensuremath{\overline q}\xspace}
\def\qqbar {\ensuremath{q\overline q}\xspace}

\def\ccbar {\ensuremath{c\overline c}\xspace}

\def\pipi      {\ensuremath{\pi\pi}\xspace}

\def\invfb     {\ensuremath{\mbox{\,fb}^{-1}}\xspace}

\newcommand{\boldk}{\mathbf{k}}

\def\aulTh{\ensuremath{{ A_{12}^{UL}}}}
\def\aucTh{\ensuremath{{ A_{12}^{UC}}}}
\def\aul{\ensuremath{{ A_{0}^{UL}}}}
\def\auc{\ensuremath{{ A_{0}^{UC}}}}

\def\zbin{\ensuremath{(z_1,z_2)}}

\def\Sq        {\mbox{$\mathbf{S}_q$}\xspace}

\mathchardef\Upsilon="7107
\def\Y#1S{\ensuremath{\Upsilon{(#1S)}}\xspace}

\def\Bbar    {\kern 0.18em\overline{\kern -0.18em B}{}\xspace}

\def\BB      {\ensuremath{B\Bbar}\xspace}

 \def\uds   {\ensuremath{uds}\xspace}
 
\def\phiz              {\ensuremath{\phi_{0}}\xspace}
\def\twophiz        {\ensuremath{2\phi_{0}}\xspace}
\def\phione         {\ensuremath{\phi_{1}}\xspace}
\def\phitwo         {\ensuremath{\phi_{2}}\xspace}
\def\phionetwo   {\ensuremath{\phi_{12}}\xspace}

\def\thrust{\ensuremath{\hat{n}}\xspace}

\def\thetath{\ensuremath{\theta_{\rm{th}}}\xspace}

\def\jetset     {\mbox{\tt Jetset}\xspace}

\def\geant      {\mbox{\sc Geant4}\xspace}

\newcommand{\BABARPubYear}    {15}
\newcommand{\BABARPubNumber}  {001}
\newcommand{\SLACPubNumber}   {16311}

\usepackage{relsize}
\def\babar{\mbox{\slshape B\kern-0.1em{\smaller A}\kern-0.1em
    B\kern-0.1em{\smaller A\kern-0.2em R}}}

\long\def\inst#1{\par\nobreak\kern 4pt\nobreak
    {\it #1}\par\vskip 10pt plus 3pt minus 3pt}

\begin{document}

\begin{flushleft}
\vspace{-2cm}
SLAC-PUB-\SLACPubNumber \\
\babar-PUB-\BABARPubYear/\BABARPubNumber
\end{flushleft}

\title{
\Large \bf {\boldmath Collins asymmetries in inclusive  
 charged \kk and \kpi pairs produced in \epem annihilation}
} 
%% author list as of 07-Feb-2015 (289 authors)
%
\author{J.~P.~Lees}
\author{V.~Poireau}
\author{V.~Tisserand}
\affiliation{Laboratoire d'Annecy-le-Vieux de Physique des Particules (LAPP), Universit\'e de Savoie, CNRS/IN2P3,  F-74941 Annecy-Le-Vieux, France}
\author{E.~Grauges}
\affiliation{Universitat de Barcelona, Facultat de Fisica, Departament ECM, E-08028 Barcelona, Spain }
\author{A.~Palano$^{ab}$ }
\affiliation{INFN Sezione di Bari$^{a}$; Dipartimento di Fisica, Universit\`a di Bari$^{b}$, I-70126 Bari, Italy }
\author{G.~Eigen}
\author{B.~Stugu}
\affiliation{University of Bergen, Institute of Physics, N-5007 Bergen, Norway }
\author{D.~N.~Brown}
\author{L.~T.~Kerth}
\author{Yu.~G.~Kolomensky}
\author{M.~J.~Lee}
\author{G.~Lynch}
\affiliation{Lawrence Berkeley National Laboratory and University of California, Berkeley, California 94720, USA }
% ...
\collaboration{The \babar\ Collaboration}
\noaffiliation

\begin{abstract}
\noindent
We present measurements of Collins asymmetries in the inclusive
process $\epem\to h_{1} h_{2} X$, $h_{1}h_{2}=\kk,\,\kpi,\,\pipi$,
at the center-of-mass energy of 10.6 GeV, using
a data sample of 468 \invfb collected by the \babar\ experiment at the PEP-II
$B$ factory at SLAC National Accelerator Center.
Considering hadrons in opposite thrust hemispheres of hadronic events, 
we observe clear azimuthal asymmetries in the ratio of unlike-sign to like-sign, 
and unlike-sign to all charged $h_{1} h_{2}$ pairs,
which increase with hadron energies.
The \kpi asymmetries are similar to those measured for the \pipi pairs,
whereas those measured for high-energy \kk pairs are, in general, larger.
\end{abstract}


\maketitle
\noindent

The Collins effect~\cite{Collins:1992kk} relates the transverse spin component of a fragmenting quark
to the azimuthal distribution of final state hadrons  about its flight direction.
The chiral-odd, transverse momentum-dependent Collins fragmentation function (FF)
provides a unique probe of quantum chromodynamics (QCD),
such as factorization and evolution with
the energy scale $Q^2$~\cite{Collins1981381,Collins1985199,PhysRevLett.93.252001,PhysRevD.88.114012}.

Additional interest has been sparked by the observation of azimuthal asymmetries
for pions and kaons in 
semi-inclusive deep inelastic scattering experiments (SIDIS) \cite{PhysRevLett.94.012002, Airapetian201011,Adolph2012376,Alekseev2009127,PhysRevD.87.012010}.
These are sensitive to the product of a Collins FF and a chiral-odd 
transversity parton distribution function (PDF), one of the three
fundamental PDFs needed to describe the spin content of the nucleon.
Although these observations require nonzero Collins FFs,
independent direct measurements of one of these chiral-odd functions are needed to determine each of them.

In \epem annihilation, one can measure the product of two Collins FFs, and
detailed measurements have been made for pairs of charged pions ~\cite{mycollins,belle1,belle2}.
No measurements are available for \kpi and \kk pairs, which are sensitive 
to different quark-flavor combinations, in particular the contribution 
 of the strange quark.
 Such measurements could be combined with SIDIS data to simultaneously 
 determine the Collins FFs and transversity PDF for up, down, and strange quarks
 ~\cite{Bacchetta2008234,PhysRevD.75.054032,PhysRevD.87.094019,PhysRevLett.67.552,PhysRevLett.74.1292,PhysRevD.91.014034}.

In this paper, we report the measurement of the Collins effect (or Collins asymmetry) for inclusive
production of  hadron pairs in the process $\epem\to\qqbar\to h_{1}
h_{2} X$, where $h_{1,2}=K^\pm$ or $\pi^\pm$, 
$q$ stands for light quarks $u$ or $d$ or $s$, and $X$ for any combination of additional hadrons.

The probability that a transversely polarized quark ($q^\uparrow$)
with momentum direction $\hat{{\boldk}}$  and spin  $\Sq$ 
 fragments into a hadron $h$ carrying zero intrinsic spin with momentum $\mathbf{P}_h$,
is defined in terms of unpolarized $ D_1^q$ and Collins $H_1^{\perp q}$  fragmentation functions~\cite{PhysRevD.70.117504}: 
\begin{equation}\label{eqn:Ndensity}
D_{h}^{q^\uparrow}(z,\mathbf{P}_{hT}) = D_1^q (z, P_{hT}^2) \\
+ H_1^{\perp q}(z,P_{hT}^2) \frac{(\hat{{\boldk}} \times \mathbf{P}_{hT})\cdot \Sq}{zM_h},
\end{equation} 
where $M_h$, $\mathbf{P}_{hT}$, and $z=2E_h/\sqrt{s}$ are the hadron mass,
momentum transverse to $\hat{\boldk}$, and fractional energy, respectively,
with $E_h$ its total energy and $\sqrt{s}$ the \epem center-of-mass (c.m.) energy.
The term including $H_1^\perp$ introduces an azimuthal modulation around the 
direction of the fragmenting quark, called Collins asymmetry.

\begin{figure}[!htb]
 \begin{center}
  \includegraphics[width=0.48\textwidth]{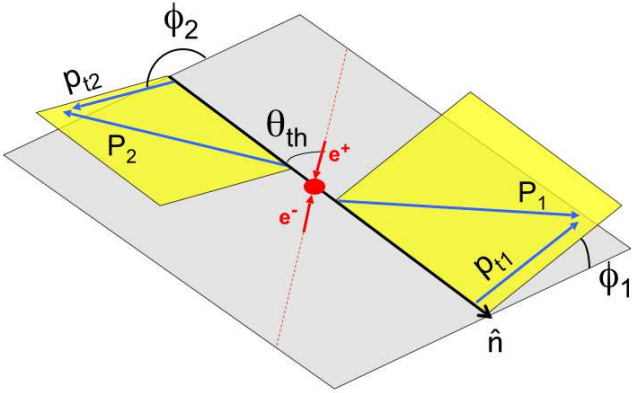}
\caption{ (color online).
Thrust reference frame (RF12).
The azimuthal angles $\phi_1$ and $\phi_2$ are the
angles between the scattering plane and the
transverse hadron momenta $\mathbf{p}_{t1(t2)}$ around the
thrust axis $\bf{\hat{n}}$.
The polar angle \thetath is the angle between $\bf{\hat{n}}$ 
and the beam axis.
Note that  the difference between $\mathbf{p}_{t1(t2)}$ and $\mathbf{P}_{hT}$ is that the latter
is calculated with respect to the \qqbar axis.
}
\label{fig:sdr12}
\end{center}
\end{figure}

In $\epem\to\qqbar$ events, the $q$ and $\qbar$ must be produced back-to-back in the \epem
c.m. frame with their spin aligned.
For unpolarized $e^+$ and $e^-$ beams at \babar\ energies, the $q$ and $\qbar$ spins are polarized along
either the $e^+$ or $e^-$ beam direction, so there is a large transverse
component when the angle between the $\epem$ and the \qqbar axis is large.
The direction is unknown for a given event, but the correlation can be exploited.
Experimentally,  the $q$ and $\qbar$ directions are difficult to measure, but the
event thrust axis \thrust~\cite{PhysRevLett.39.1587,Brandt196457} approximates
at leading order the \qqbar axis, so an azimuthal correlation between two hadrons in opposite thrust hemispheres reflects
the product of the two Collins functions.

Figure \ref{fig:sdr12} shows the thrust reference frame (RF12)~\cite{Daniel200923}.
If not otherwise specified, all kinematic variables are defined in the \epem c.m. frame.
The Collins effect results in a cosine modulation of the azimuthal angle $\phionetwo=\phione+\phitwo$
of the di-hadron yields. 
Expressing the yield as a function of \phionetwo (after the integration over $\mathbf{P}_{hT}$), 
and dividing by the average bin content, we obtain the normalized rate~\cite{mycollins} 
\begin{equation}\label{eq:r12}
R_{12}(\phionetwo)= 1+ \frac{\sin^2\thetath}{ 1+\cos^2\thetath} \cos\phionetwo \cdot \frac{H_1^{\perp[1]}(z_1) \overline{H}_1^{\perp[1]}(z_2)}{D_1^{[0]}(z_1) \overline{D}^{[0]}_1(z_2)},
\end{equation}
where the sum over the involved quark flavors is implied,
 \thetath is defined in Fig. \ref{fig:sdr12},  $z_{1(2)}$ is the fractional energy of the first (second) hadron,
and the bar denotes the function for the \qbar.
Equation~(\ref{eq:r12}) involves only the moments of FF, which are defined as
\begin{equation}
F^{[n]}(z_i)\equiv \int d|\mathbf{k_T}|^2 \left[ \frac{|\mathbf{k_T}|}{M_i} \right]^n F (z_i, |\mathbf{k_T}|^2),
\end{equation}
with $n=0,\, 1$,  
and $|\mathbf{k_T}|$ the transverse momentum of the quarks with respect to the hadrons they fragment into,
which, in this frame, is related to the measurement of the transverse momenta of the two hadrons with respect
to the thrust axis.
 
Despite the simple form of the $R_{12}$ normalized rate, which involves only the product of moments of FFs,
the RF12 frame comes with several downsides, among others of having to rely 
on Monte Carlo (MC) simulations when using the thrust axis as a proxy for the leading-order \qqbar axis.
An alternative frame is the analogue of the Gottfried-Jackson frame~\cite{Boer1997345,Daniel200923} which
uses the momentum of one hadron as a reference axis, and defines
a single angle \phiz between the plane containing the two hadron momenta
and the plane defined by the beam and the reference axis.
We refer to this frame as RF0~\cite{mycollins,belle1}.
The corresponding normalized yield in the \epem c.m. system is~\cite{Daniel200923}
\begin{equation}
\begin{split}
R_{0}(\twophiz) = &1+ \frac{\sin^2\theta_2}{ 1+\cos^2\theta_2}\cos\twophiz \cdot \\
& \frac{\mathcal{F}[(2\mathbf{\hat{h}}\cdot \mathbf{k}_T \, \mathbf{\hat{h}}\cdot \mathbf{p}_T - \mathbf{k}_T\cdot \mathbf{p}_T)
H_1^{\perp} \overline{H}_1^{\perp}]}{(M_1M_2) \mathcal{F}[D_1 \overline{D}_1]},
\end{split}
\end{equation}
where  $\theta_2$ is the angle between the hadron used as reference and the beam axis,
$\mathbf{\hat{h}}$ is the unit vector in the direction of the transverse momentum
of the first hadron relative to the axis defined by the second hadron,
and $\mathcal{F}$ is used to denote the convolution integral
\begin{equation}
\begin{split}
\mathcal{F}[X\overline{X}]\equiv & \sum_{q} e_q^2 \int d^2 \mathbf{k}_T  d^2 \mathbf{p}_T \delta^2 ( \mathbf{p}_T+\mathbf{k}_T -\mathbf{q}_T) \\
& X^q(z_1,z_1^2\mathbf{k}_T^2) \overline{X}^q(z_2,z_2^2\mathbf{p}_T^2), 
\end{split}
\end{equation}
with $\mathbf{k}_T$, $\mathbf{p}_T$, and $\mathbf{q}_T$ the transverse momentum
of the fragmenting quark,  antiquark, and virtual photon from \epem annihilation, respectively,
in the frame where the two hadrons are collinear, and $X\,(\overline{X})\equiv D_1\, (\overline{D}_1)$ or $H^\perp_1\, (\overline{H}^\perp_1)$.
In this frame, specific assumptions on the $\mathbf{k}_T$-dependence of the involved functions
are necessary to explicitly evaluate the convolution integrals.

For this analysis 
we use a data sample of 468 \invfb \cite{Lees:2013rw} collected at the c.m. energy $\sqrt{s}\approx 10.6$ \gev
with the \babar\ detector \cite{Aubert20021,Aubert2013615} at the SLAC National Accelerator Laboratory.
We use tracks reconstructed in the silicon vertex detector and in the drift chamber (DCH)
and identified as pions or kaons in the DCH and in the Cherenkov ring imaging detector (DIRC).
Detailed MC simulation is used to study detector effects and to estimate
contribution from various background sources.
Hadronic events are generated using the \jetset \cite{Sjostrand:1995iq} package and undergo 
a full detector simulation based on \geant \cite{Agostinelli:2002hh}.

We make a tight selection of hadronic events in order to minimize biases
due to detector acceptance and hard initial-state photon radiation (ISR),
as they can introduce fake azimuthal modulations.
Furthermore, final-state gluon ($\qqbar g$) radiation also leads to angular asymmetries
to be taken into account~\cite{Daniel200923}.
Requiring at least three charged tracks consistent with 
the $e^+e^-$ primary vertex and a total visible energy of the event in the laboratory frame $E_{\rm{tot}}>11$ GeV,
we reject $\epem\rightarrow\tautau$  and two-photon backgrounds, as well as 
ISR ($\qqbar g$) events with the photon (one jet) along the beam line.
About $10\%$ of ISR photons are within our detector acceptance,
and we reject events with a photon candidate with energy above 2 GeV.
We require an event thrust value $T>0.8$ to suppress $\qqbar g$ and \BB
events, and $|\cos\thetath|<0.6$ so that most tracks are within the detector acceptance.

We assign randomly the positive direction of the thrust axis, and divide
each event into two hemispheres by the plane perpendicular to it.
To ensure tracks are assigned to the correct hemispheres,
we require them to be within a $45^\circ$ angle of the thrust axis and to have $z>0.15$.
A ``tight" identification algorithm is used to identify kaons (pions), which is about $80\%$ ($90\%$) efficient 
and has misidentification rates below $10\%$ ($5\%$).
We select those pions and kaons that lie within the DIRC acceptance region with 
a polar angle in laboratory frame $0.45\,\rm{rad}<\theta_{\rm{lab}}<2.46\,\rm{rad}$.
To minimize backgrounds, such as $\epem\to\mu^+\mu^-\gamma$ followed by photon conversion, we require $z<0.9$.

We construct all the possible pairs of selected tracks reconstructed in opposite thrust hemispheres,
and we calculate the corresponding
azimuthal angles \phione, \phitwo, and \phiz in the respective
reference frames.
In this way, we identify three different samples of hadron pairs: \kk, \kpi, and \pipi. 
To reduce low-energy gluon radiation and the contribution due to wrong hemispheres assignment,
we require $Q_t<3.5$ \gevc, where $Q_t$ is the transverse momentum of the virtual photon from
\epem annihilation in the frame where the two hadrons are collinear~\cite{Daniel200923}.

The analysis is performed in intervals of hadron fractional energies with 
the following boundaries: 0.15, 0.2, 0.3, 0.5, 0.9,  for a total of 16
two-dimensional \zbin\ intervals. 

\begin{figure}[!htb]
 \begin{center}
  \includegraphics[width=0.48\textwidth]{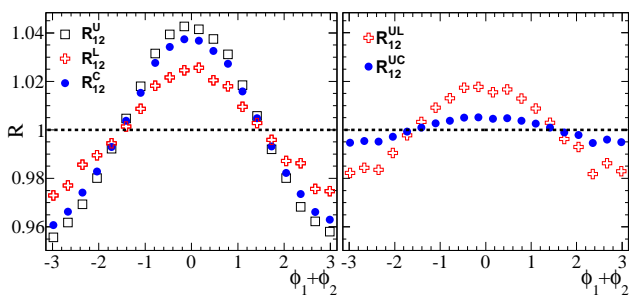}
\caption{ (color online). 
 Distributions of normalized yields (left plot) for unlike ($U$), like ($L$), and any
charge combination ($C$) of KK pairs, and their double ratios (right plot) in RF12. 
}
\label{fig:distr}
\end{center}
\end{figure}

For each of the three samples, we evaluate the normalized yield distributions $R_{12}$ and $R_{0}$
for unlike ($U$), like ($L$), and any charge combination ($C$) of hadron pairs
as a function of $\phi_1+\phi_2$ and $2\phi_0$,
as shown in the left plot of Fig.~\ref{fig:distr} for $KK$ pairs, for example.
These combinations of charged hadrons contain different contributions of favored and 
disfavored FFs, where a favored (disfavored) process refers to the
production of a hadron for which one (none) of the valence quarks is of the same kind as the
fragmenting quark.
In particular, by selecting \kk pairs, we are able to study the favored contribution $H_s^{\perp \rm{fav}}$
of the strange quark, not accessible when considering \pipi pairs only.

The normalized distributions can be  parametrized with a cosine function: 
$R_{\alpha}^i=b_\alpha+a^i_\alpha\cos\beta_\alpha$,
where $\alpha=0,\,12$ indicates the reference frames, 
$i=U,\,L,\,C$ the charge combination of hadron pairs,
and $\beta_{12(0)}=\phionetwo (\twophiz)$.

The $R_\alpha^i$ distributions are strongly affected by instrumental effects.
In order to reduce the impact of the detector acceptance,
as well as any remaining effect from gluon bremsstrahlung~\cite{Daniel200923}, we construct two
double ratios (DR) of normalized distributions, $R^U_\alpha/R^L_\alpha$ and $R^U_\alpha/R^C_\alpha$.
The two ratios give access to the same physical quantities as the 
independent  $R_\alpha^i$, that is the favored and disfavored FFs, but in different combinations.
We report the results for both kind of DRs,
which are strongly correlated since they are obtained by using the same data set. 
These are shown in the right plot of Fig.~\ref{fig:distr} for \kk pairs in RF12.
At first order, the double ratios are still parametrized by a 
function that is linear in the cosine of the corresponding combination of azimuthal angles:
\begin{equation}\label{eq:cos}
R^{ij}_\alpha=\frac{R_{\alpha}^{i}}{R_\alpha^{j}} \simeq
B^{ij}_\alpha+A^{ij}_\alpha \cdot \cos\beta_\alpha,
\end{equation}
with $B$ and $A$ free parameters, and $i,\,j=U,\,L,\,C$.
The constant term $B$ must be consistent with unity, while $A$ contains the information
about the favored and disfavored Collins FFs.

We fit the binned $R^{ij}_\alpha$ distributions independently for \kk, \kpi, and \pipi hadron pairs.
Using the MC sample, we evaluate the $K/\pi$ (mis)identification probabilities for
the 16 \zbin\ intervals in each of the three samples. For example,
the probability $f_{\kk}^{\kk}$ that a true \kk pair is reconstructed as \kk pair is about 90\% on average,
slightly decreasing at higher momenta,
while the probability $f_{\kpi}^{\kk}$ that a true \kpi pair is identified as \kk is 
about 10\%, and $f_{\pipi}^{\kk}$ is negligible.

The presence of background processes could introduce azimuthal modulations
not related to the Collins effect, and modifies the measured asymmetry as follows:
\begin{equation}\label{eq:iter1}
\begin{split}
A^{\rm{meas}}_{\kk}=& F_{uds}^{\kk}  \cdot \left( \sum_{nm} f_{nm}^{KK} \cdot A_{nm} \right)  +  \\
 & \sum_i F_{i}^{KK} \left( \sum_{nm} f_{nm}^{(KK)i} \cdot  A^{i}_{nm}  \right),
\end{split}
\end{equation}
with $nm=\kk,\,\kpi,\,\pipi$, and $i= \ccbar,\, \BB,\, \tautau$.
In Eq.~\ref{eq:iter1}, $A_{nm}$ are the true Collins asymmetries produced from the
fragmentation of light quarks in the three samples,
$A^{i}_{nm}$ is the $i$-th background asymmetry contribution,
and $F_{uds (i)}^{KK}$ are the fractions of reconstructed kaon pairs coming from $uds$ and background events,
calculated from the respective MC samples.
By construction, $\sum_i F_{i}+F_{uds}=1$. 
A similar expression holds for \kpi and \pipi samples.

Previous studies~\cite{mycollins} show that $\epem\to\BB$ and \tautau events
have negligible $A_{nm}^{i}$,
$F_{\BB}<2\%$, and $F_{\tautau}$ significantly different from zero
only for the \pipi sample at high $z$ values.
Since $F_{\ccbar}$ can be as large as 30\%, and $A^{\ccbar}$ are unknown,
we determine $A^{\ccbar}_{nm}$ in Eq.~(\ref{eq:iter1}) from samples enhanced in $\ccbar$
by requiring the reconstruction of at least one $D^{*\pm}$ meson
from the decay $D^{*\pm}\to D^0\pi^\pm$, with the $D^0$ candidate 
reconstructed in the following four Cabibbo-favored decay modes: $K^-\pi^+$,
$K^-\pi^+\pi^-\pi^+$, $K^0_s\pi^+\pi^-$, and $K^-\pi^+\pi^0$.
These modes are assumed to provide a representative sample of \pipi, \kpi, and \kk
pairs to be used in the correction, an assumption that is strengthened by the observation
that the background asymmetries for those modes were found to be consistent.
We solve the system of equations for $A^{\rm{meas}}_{KK},\,
A^{\rm{meas}}_{\kpi}$, $A^{\rm{meas}}_{\pipi}$,  for the standard and
charm-enhanced samples, and we extract simultaneously the Collins
asymmetries $A_{KK},\, A_{\kpi}$, and $A_{\pipi}$, corrected  
for the contributions of the background and $K/\pi$ (mis)identification.
The dominant uncertainties related to this procedure come from 
the limited statistics of the $D^*$-enhanced sample and from the 
fractions $F_i$. The uncertainties on the fractions are evaluated by data-MC comparison
and amount to a few percent.
All these uncertainties are therefore included in the statistical error of the asymmetries
extracted from the system of Eq.~(\ref{eq:iter1}).

We test the DR method on the MC sample.
Spin effects are not simulated in MC, and so the DR distributions should be uniform.
However, when fitting the distributions for reconstructed \kk pairs with Eq.~(\ref{eq:cos}),
we measure a cosine term in the full sample of 
$0.004\pm0.001$ and $0.007\pm0.001$ in the RF12 and RF0 frames, respectively, indicating a bias.
Smaller values are obtained for \kpi and \pipi pairs (see Fig.~\ref{fig:bias}).
Studies performed on the MC samples, both at generation level and
after full simulation, demonstrate that the main source of this bias
is due to the emission of ISR, which boosts the hadronic system and distorts
the angular distribution of the final state particles,
resulting in azimuthal modulations not related to the Collins effect.
This effect is more pronounced for $KK$ pairs due to the lower multiplicity
with respect to the other two combinations of hadrons.
Assuming the bias, which is everywhere smaller than the asymmetries
measured in the data sample in each bin, is additive, we subtract it from the background-corrected asymmetry.

\begin{figure}[!ht]
\centering
\includegraphics[scale=0.43]{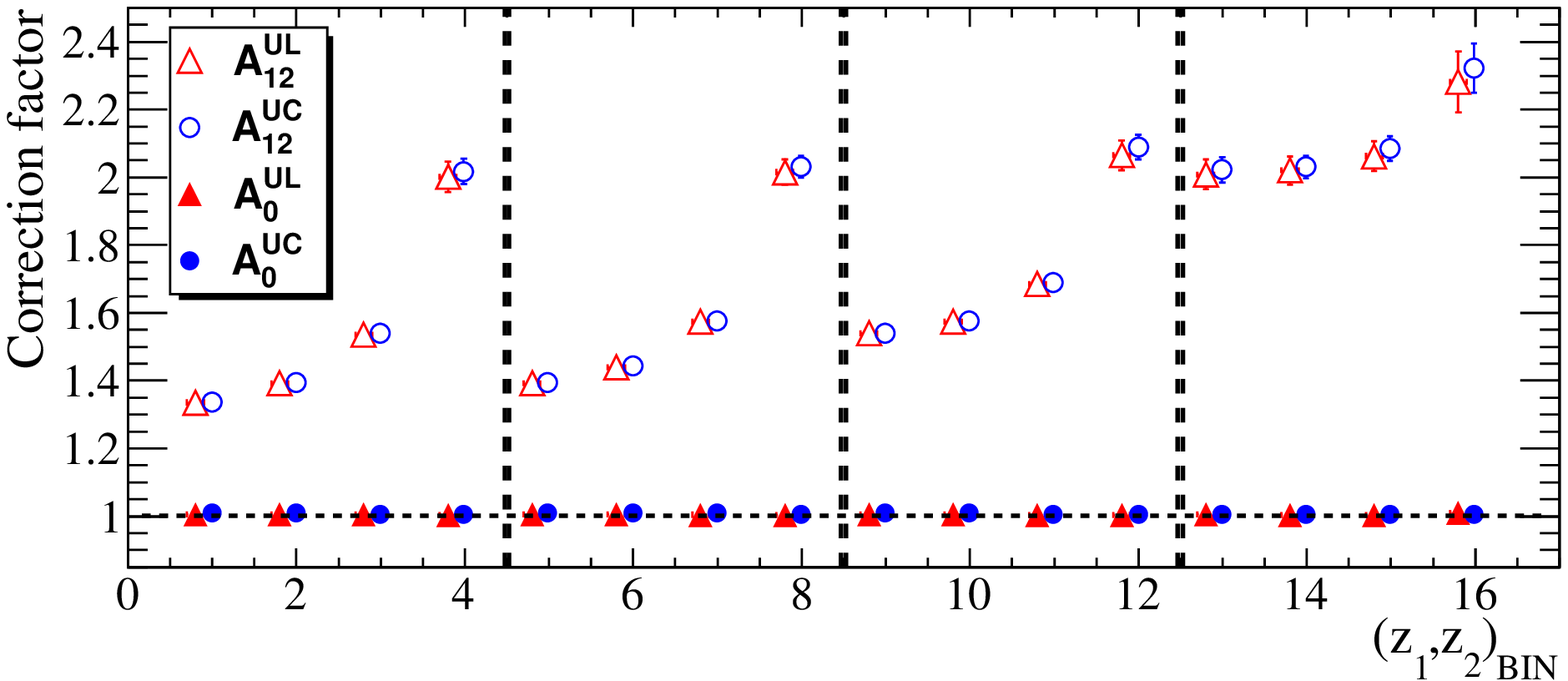}\\
\includegraphics[scale=0.43]{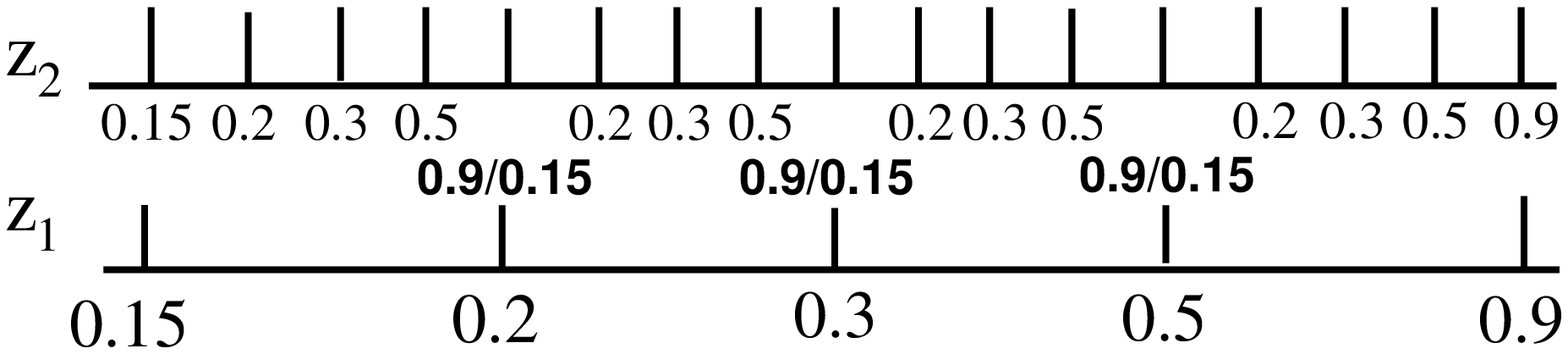}
\caption{ (color online).
Correction factors for the dilution of the asymmetry due to the difference between the thrust and the $q\bar{q}$ axis.
The open (full) markers, triangles and circles, refer to the U/L
and U/C double ratios in the RF12 (RF0) frame, respectively.
The 16 ($z_1,z_2$) bins are shown on the x-axis:
in each interval between the dashed lines, $z_1$ is chosen in the
following ranges: $[0.15,0.2]$, $[0.2,0.3]$, $[0.3,0.5]$, and $[0.5,0.9]$, while
within each interval the points correspond to the four bins in $z_2$.
}
\label{fig:thrustCorr}
\end{figure}

Using the $uds$ MC sample, or light quark $\epem\to\qqbar$ MC events,
we study the difference between measured and true azimuthal asymmetries.
The asymmetry is introduced into the simulation by reweighting the events according to
the distribution $1\pm a\cdot \cos\phi_{\alpha}^{\rm{gen}}$, where we use 
different values of $a$ ranging from $0$ to $8\%$
with positive (negative) sign for $U$ ($L$ and $C$ ) hadron pairs, and 
$\phi_{\alpha}^{\rm{gen}}$ are the azimuthal angles combinations calculated with respect to the true \qqbar
axis in RF12, or the generated hadron momentum in RF0.
The reconstructed asymmetries in RF12 are systematically underestimated for the three
samples of hadron pairs, as expected since we use the thrust axis instead of the \qqbar axis,
while they are consistent with the simulated ones in RF0, where
only particle identification and tracking reconstruction effects could introduce possible dilution.
Since we measure the same dilution for \kk, \kpi, and \pipi samples,  
the asymmetry is corrected by rescaling \akk, \akpi, and \apipi using the same
correction factor, which ranges from 1.3 to 2.3 increasing with $z$,
as shown in Fig.~\ref{fig:thrustCorr}.
No corrections are needed for the asymmetries measured in RF0.
The uncertainties on the correction factors are assigned 
as systematic contributions.

\begin{figure*}[!ht]
\begin{minipage}[b]{0.48\linewidth}
\centering
\includegraphics[width=\textwidth]{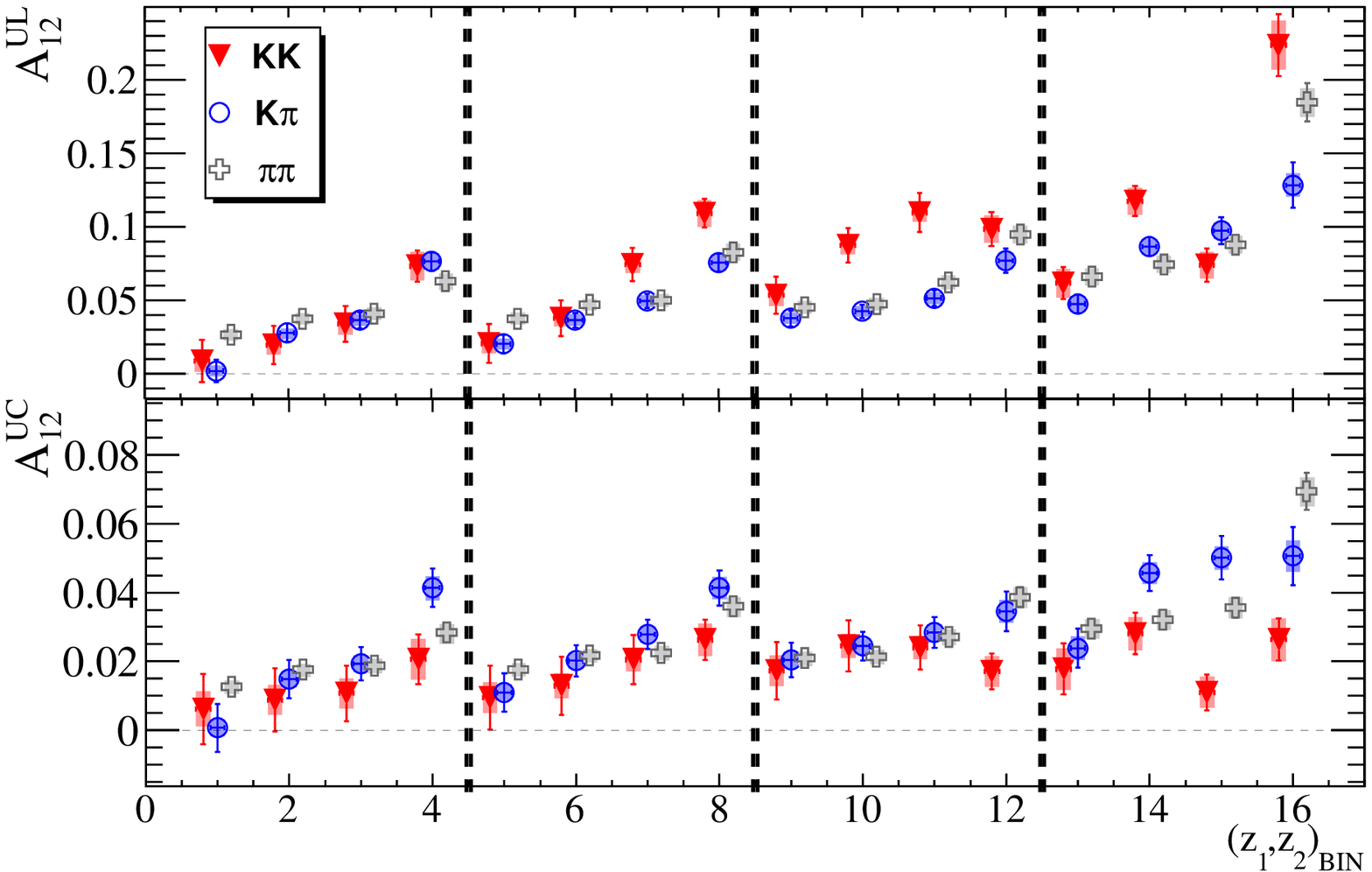}\\
\includegraphics[width=\textwidth]{doublebins}
\end{minipage}
\hspace{0.5cm}
\begin{minipage}[b]{0.48\linewidth}
\centering
\includegraphics[width=\textwidth]{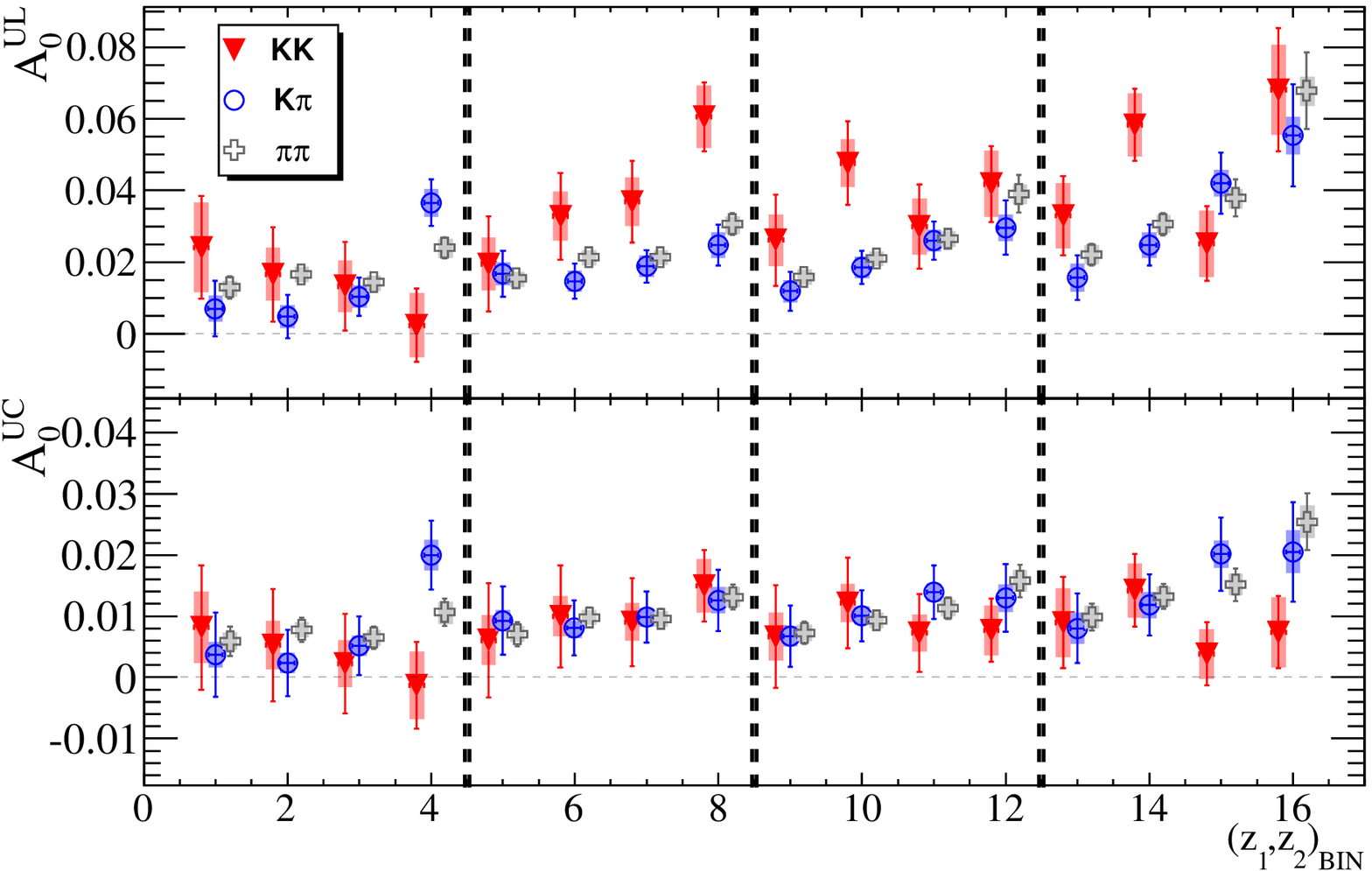}
\includegraphics[width=\textwidth]{doublebins}
\end{minipage}
\caption{ (color online). Comparison of U/L (top) and U/C (bottom) Collins asymmetries in RF12 (left)
and RF0 (right) for \kk, \kpi, and \pipi pairs.
The statistical and systematic uncertainties are represented by the bars and the bands around the points, respectively.
The 16 \zbin\ bins are shown on the x-axis:
in each interval between the dashed lines, $z_1$ is chosen in the
following ranges: $[0.15,0.2]$, $[0.2,0.3]$, $[0.3,0.5]$, and $[0.5,0.9]$, while
within each interval the points correspond to the four bins in $z_2$.
}
\label{fig:rf120}
\end{figure*}

All systematic effects, if not otherwise specified, are evaluated for
each bin of $z$.
The main contribution comes from the MC bias.
We compare the bias results from the nominal selection,
with those obtained by requiring different cuts on $E_{\rm{tot}}$, and/or by changing the detector  
acceptance region for the hadrons.
The largest variation of the bias is combined in quadrature with the 
MC statistical error and taken as systematic uncertainty.
The effects due to the particle identification are evaluated using tighter and looser
selection criteria.  The largest deviations with respect to
the nominal selection are taken as systematic uncertainties: 
the average relative uncertainties are around $10\%$,
$7\%$, and $5\%$ for the \kk, \kpi, and \pipi pairs. 
Fitting the azimuthal distributions using different bin sizes, we
determine relative systematic uncertainties, which are not larger than $5\%$, $1.9\%$, and
$1\%$ for the three samples.
The systematic uncertainty due to the $E_{\rm{tot}}$ cut is obtained by comparing the measured asymmetries 
with those obtained with the looser selection $E_{\rm{tot}}>10$ GeV.
The average systematic contribution is around $10\%$ for the three samples
in both reference frames.
We use different fitting functions with additional higher harmonic terms.
No significant changes in the value of the cosine moments
with respect to the standard fits are found.
As a cross-check of the double ratio method we fit the difference 
of $R^i$ distributions, and we compare the two results.
The difference between the two procedures is negligible for \kpi and \pipi pairs,
while it reaches $1\%$ and $3\%$ for kaon pairs in RF12 and RF0, respectively.
All the other systematic contributions are negligible~\cite{mycollins}.

The Collins asymmetries measured for the 16 two-dimensional
$(z_1,z_2)$ bins, for reconstructed \kk, \kpi, and \pipi hadron pairs,
are shown in Fig.\,\ref{fig:rf120} for RF12 and RF0,
and are summarized in Tables~\ref{tab:KKtotZ},~\ref{tab:KPitotZ}, and~\ref{tab:PiPitotZ} . 
The asymmetries are corrected for the background contributions and 
$K/\pi$ contamination following Eq.~(\ref{eq:iter1}), 
the MC bias is subtracted, and the corrections due to the dilution effects are applied.
The total systematic uncertainties are obtained by adding in quadrature 
the individual contributions, and are represented by the bands around the data points.

An increasing asymmetry with increasing hadron energies is visible
for the $U/L$ double ratio in both reference frames.
The largest effects, but with less precision, are observed for \kk pairs, for which
$A^{UL}_{12}$ is consistent with zero
at low $z$, and reaches $22\%$ in the last $z$ bin, while somewhat
smaller values are seen for \pipi and \kpi pairs.
In particular, at low \zbin\ bins $A^{UL}$ for \pipi pairs is nonzero, in agreement 
with the behavior observed in~\cite{mycollins}.
The small differences between the two data sets are due to the
different kinematic region selected after the cut on $\cos\theta_{th}$.
The $A^{UC}$ asymmetry is smaller than $A^{UL}$ in all cases,
and, for the \kk pairs, the rise of the asymmetry with the hadron energies is
not evident.
We also note that the asymmetries for the \kk pairs are larger than the others when the
U/L ratio is considered, while they are at the same level, or lower, when they are
extracted from the U/C ratio.

In summary, we have studied for the first time in \epem annihilation
the Collins asymmetry for inclusive production of \kk and \kpi pairs as a function of
\zbin\ in two distinct reference frames.
We measure the azimuthal modulation of the
double ratios $U/L$ and $U/C$, 
which are sensitive to the favored and disfavored Collins FFs for light quarks.
We simultaneously extract also 
the Collins asymmetries for \pipi pairs, which are found to be in agreement
with those obtained in previous studies\,\cite{mycollins,belle2}.
The results reported in this paper and those obtained from 
SIDIS experiments can be used in a global analysis
to extract the favored contribution of the strange quark, and to improve the
knowledge on the $u$ and $d$ fragmentation processes~\cite{PhysRevD.75.054032,PhysRevD.87.094019,Bacchetta2008234}.\\

We are grateful for the excellent luminosity and machine conditions
provided by our \pep2\ colleagues,
and for the substantial dedicated effort from
the computing organizations that support \babar.
The collaborating institutions wish to thank
SLAC for its support and kind hospitality.
This work is supported by
DOE
and NSF (USA),
NSERC (Canada),
CEA and
CNRS-IN2P3
(France),
BMBF and DFG
(Germany),
INFN (Italy),
FOM (The Netherlands),
NFR (Norway),
MES (Russia),
MEC (Spain), and
STFC (United Kingdom).
Individuals have received support from the
Marie Curie EIF (European Union) and
the A.~P.~Sloan Foundation.

\bibliography{mybiblio}{}

\appendix

\begin{table*}[!hbt]
\begin{center}
\begin{tabular}{ c c c c c c c}
\toprule 
\multicolumn{7}{c}{\kk sample}\\
\toprule 
$z_1$ & $\langle z_1\rangle$ & $z_2$ & $\langle z_2\rangle$ & $  \frac{\langle\sin^2\theta_{th}\rangle}{\langle1+\cos^2\theta_{th}\rangle}$ & \aulTh\ ($10^{-2}$) & \aucTh\ ($10^{-2}$) \\ \toprule
$[0.15,0.2]$ & 0.175 & $[0.15,0.2]$ & 0.175 & 0.797 & 0.88 $\pm$ 1.43 $\pm$ 0.73 & 0.62 $\pm$ 1.02 $\pm$ 0.51\\
 $[0.15,0.2]$ & 0.175 & $[0.2,0.3]$ & 0.247 & 0.794 & 1.96 $\pm$ 1.32 $\pm$ 0.67 & 0.88 $\pm$ 0.92 $\pm$ 0.43\\
 $[0.15,0.2]$ & 0.175 & $[0.3,0.5]$ & 0.381 & 0.794 & 3.38 $\pm$ 1.23 $\pm$ 0.73 & 1.08 $\pm$ 0.81 $\pm$ 0.43\\
 $[0.15,0.2]$ & 0.175 & $[0.5,0.9]$ & 0.608 & 0.786 & 7.32 $\pm$ 1.06 $\pm$ 0.97 & 2.06 $\pm$ 0.72 $\pm$ 0.59\\
 $[0.2,0.3]$ & 0.246 & $[0.15,0.2]$ & 0.175 & 0.794 & 2.06 $\pm$ 1.33 $\pm$ 0.68 & 0.95 $\pm$ 0.93 $\pm$ 0.44\\
 $[0.2,0.3]$ & 0.247 & $[0.2,0.3]$ & 0.247 & 0.792 & 3.78 $\pm$ 1.22 $\pm$ 0.58 & 1.30 $\pm$ 0.84 $\pm$ 0.36\\
 $[0.2,0.3]$ & 0.247 & $[0.3,0.5]$ & 0.382 & 0.792 & 7.44 $\pm$ 1.14 $\pm$ 0.59 & 2.06 $\pm$ 0.72 $\pm$ 0.35\\
 $[0.2,0.3]$ & 0.247 & $[0.5,0.9]$ & 0.608 & 0.783 & 10.91 $\pm$ 0.98 $\pm$ 0.91 & 2.63 $\pm$ 0.58 $\pm$ 0.48\\
 $[0.3,0.5]$ & 0.381 & $[0.15,0.2]$ & 0.175 & 0.794 & 5.34 $\pm$ 1.28 $\pm$ 0.74 & 1.73 $\pm$ 0.84 $\pm$ 0.44\\
 $[0.3,0.5]$ & 0.381 & $[0.2,0.3]$ & 0.247 & 0.792 & 8.74 $\pm$ 1.17 $\pm$ 0.59 & 2.45 $\pm$ 0.74 $\pm$ 0.35\\
 $[0.3,0.5]$ & 0.382 & $[0.3,0.5]$ & 0.382 & 0.792 & 10.97 $\pm$ 1.31 $\pm$ 0.63 & 2.41 $\pm$ 0.65 $\pm$ 0.34\\
 $[0.3,0.5]$ & 0.383 & $[0.5,0.9]$ & 0.610 & 0.784 & 9.84 $\pm$ 1.16 $\pm$ 0.92 & 1.71 $\pm$ 0.52 $\pm$ 0.45\\
 $[0.5,0.9]$ & 0.609 & $[0.15,0.2]$ & 0.175 & 0.785 & 6.15 $\pm$ 1.09 $\pm$ 0.98 & 1.78 $\pm$ 0.74 $\pm$ 0.60\\
 $[0.5,0.9]$ & 0.608 & $[0.2,0.3]$ & 0.248 & 0.783 & 11.75 $\pm$ 1.03 $\pm$ 0.91 & 2.81 $\pm$ 0.60 $\pm$ 0.48\\
 $[0.5,0.9]$ & 0.610 & $[0.3,0.5]$ & 0.383 & 0.784 & 7.40 $\pm$ 1.13 $\pm$ 0.91 & 1.11 $\pm$ 0.52 $\pm$ 0.45\\
 $[0.5,0.9]$ & 0.615 & $[0.5,0.9]$ & 0.615 & 0.776 & 22.36 $\pm$ 2.09 $\pm$ 1.69 & 2.63 $\pm$ 0.61 $\pm$ 0.62\\
\toprule
$z_1$ & $\langle z_1\rangle$ & $z_2$ & $\langle z_2\rangle$ &  $ \frac{\langle\sin^2\theta_{2}\rangle}{\langle1+\cos^2\theta_{2}\rangle}$ & \aul\ ($10^{-2}$) & \auc\ ($10^{-2}$) \\ \toprule
$[0.15,0.2]$ & 0.175 & $[0.15,0.2]$ & 0.175 & 0.739 & 2.41 $\pm$ 1.44 $\pm$ 1.25 & 0.82 $\pm$ 1.02 $\pm$ 0.58\\
 $[0.15,0.2]$ & 0.175 & $[0.2,0.3]$ & 0.247 & 0.736 & 1.66 $\pm$ 1.31 $\pm$ 0.74 & 0.53 $\pm$ 0.92 $\pm$ 0.40\\
 $[0.15,0.2]$ & 0.175 & $[0.3,0.5]$ & 0.381 & 0.750 & 1.33 $\pm$ 1.24 $\pm$ 0.71 & 0.23 $\pm$ 0.81 $\pm$ 0.39\\
 $[0.15,0.2]$ & 0.175 & $[0.5,0.9]$ & 0.608 & 0.751 & 0.24 $\pm$ 1.02 $\pm$ 0.90 & -0.13 $\pm$ 0.71 $\pm$ 0.55\\
 $[0.2,0.3]$ & 0.246 & $[0.15,0.2]$ & 0.175 & 0.739 & 1.95 $\pm$ 1.32 $\pm$ 0.75 & 0.61 $\pm$ 0.93 $\pm$ 0.41\\
 $[0.2,0.3]$ & 0.247 & $[0.2,0.3]$ & 0.247 & 0.736 & 3.28 $\pm$ 1.21 $\pm$ 0.69 & 1.00 $\pm$ 0.84 $\pm$ 0.33\\
 $[0.2,0.3]$ & 0.247 & $[0.3,0.5]$ & 0.382 & 0.749 & 3.69 $\pm$ 1.14 $\pm$ 0.67 & 0.90 $\pm$ 0.72 $\pm$ 0.31\\
 $[0.2,0.3]$ & 0.247 & $[0.5,0.9]$ & 0.608 & 0.750 & 6.05 $\pm$ 0.96 $\pm$ 0.88 & 1.49 $\pm$ 0.58 $\pm$ 0.44\\
 $[0.3,0.5]$ & 0.381 & $[0.15,0.2]$ & 0.175 & 0.738 & 2.62 $\pm$ 1.27 $\pm$ 0.72 & 0.67 $\pm$ 0.84 $\pm$ 0.39\\
 $[0.3,0.5]$ & 0.381 & $[0.2,0.3]$ & 0.247 & 0.736 & 4.76 $\pm$ 1.17 $\pm$ 0.67 & 1.21 $\pm$ 0.74 $\pm$ 0.31\\
 $[0.3,0.5]$ & 0.382 & $[0.3,0.5]$ & 0.382 & 0.749 & 2.99 $\pm$ 1.18 $\pm$ 0.78 & 0.73 $\pm$ 0.63 $\pm$ 0.31\\
 $[0.3,0.5]$ & 0.383 & $[0.5,0.9]$ & 0.610 & 0.750 & 4.18 $\pm$ 1.06 $\pm$ 0.92 & 0.77 $\pm$ 0.52 $\pm$ 0.41\\
 $[0.5,0.9]$ & 0.609 & $[0.15,0.2]$ & 0.175 & 0.731 & 3.30 $\pm$ 1.10 $\pm$ 0.91 & 0.90 $\pm$ 0.74 $\pm$ 0.56\\
 $[0.5,0.9]$ & 0.608 & $[0.2,0.3]$ & 0.248 & 0.728 & 5.83 $\pm$ 1.00 $\pm$ 0.88 & 1.42 $\pm$ 0.60 $\pm$ 0.44\\
 $[0.5,0.9]$ & 0.610 & $[0.3,0.5]$ & 0.383 & 0.743 & 2.52 $\pm$ 1.04 $\pm$ 0.92 & 0.38 $\pm$ 0.51 $\pm$ 0.41\\
 $[0.5,0.9]$ & 0.615 & $[0.5,0.9]$ & 0.615 & 0.743 & 6.81 $\pm$ 1.72 $\pm$ 1.26 & 0.74 $\pm$ 0.59 $\pm$ 0.57\\
\toprule 
\end{tabular}
\caption {Light quark (\uds) Collins asymmetries obtained by fitting the U/L and U/C double ratios
as a function of \zbin\ for kaon pairs in the RF12 frame (upper table) and in the RF0 frame (lower table).
In the first two columns are reported the  $z$ bins and their respective
mean values for the kaon in one hemisphere; in the following two columns, 
the same variables for the second kaon are shown; in the fifth column is summarized 
the ratio of mean values  $\langle\sin^2\theta_{th(2)}\rangle/\langle 1+\cos^2\theta_{th(2)}\rangle$,
and the asymmetry results are reported in the last two columns.
The quoted errors are statistical and systematic, respectively.
The mean values of the quantities reported in the table are calculated by summing the corresponding values for
each \kk pair and dividing by the number of \kk pairs that fall into each \zbin\ interval.
Note that the $A^{UL}$ and $A^{UC}$ results are strongly correlated since they are obtained
by using the same data set.}
\label{tab:KKtotZ}
\end{center}
\end{table*}

\begin{table*}[!hbt]
\begin{center}
\begin{tabular}{ c c c c c c c}
\toprule 
\multicolumn{7}{c}{\kpi sample}\\
\toprule 
$z_1$ & $\langle z_1\rangle$ & $z_2$ & $\langle z_2\rangle$ & $ \frac{\langle\sin^2\theta_{th}\rangle}{\langle1+\cos^2\theta_{th}\rangle}$ & \aulTh\ ($10^{-2}$) & \aucTh\ ($10^{-2}$) \\ \toprule
$[0.15,0.2]$ & 0.174 & $[0.15,0.2]$ & 0.174 & 0.794 & 0.19 $\pm$ 0.77 $\pm$ 0.37 & 0.07 $\pm$ 0.69 $\pm$ 0.25\\
 $[0.15,0.2]$ & 0.174 & $[0.2,0.3]$ & 0.246 & 0.792 & 2.73 $\pm$ 0.62 $\pm$ 0.36 & 1.49 $\pm$ 0.55 $\pm$ 0.23\\
 $[0.15,0.2]$ & 0.174 & $[0.3,0.5]$ & 0.380 & 0.791 & 3.64 $\pm$ 0.53 $\pm$ 0.39 & 1.93 $\pm$ 0.48 $\pm$ 0.24\\
 $[0.15,0.2]$ & 0.174 & $[0.5,0.9]$ & 0.612 & 0.784 & 7.63 $\pm$ 0.64 $\pm$ 0.56 & 4.12 $\pm$ 0.56 $\pm$ 0.35\\
 $[0.2,0.3]$ & 0.246 & $[0.15,0.2]$ & 0.174 & 0.791 & 2.02 $\pm$ 0.63 $\pm$ 0.36 & 1.10 $\pm$ 0.56 $\pm$ 0.23\\
 $[0.2,0.3]$ & 0.245 & $[0.2,0.3]$ & 0.246 & 0.790 & 3.64 $\pm$ 0.50 $\pm$ 0.38 & 2.02 $\pm$ 0.45 $\pm$ 0.23\\
 $[0.2,0.3]$ & 0.245 & $[0.3,0.5]$ & 0.380 & 0.789 & 4.94 $\pm$ 0.47 $\pm$ 0.39 & 2.79 $\pm$ 0.42 $\pm$ 0.24\\
 $[0.2,0.3]$ & 0.245 & $[0.5,0.9]$ & 0.611 & 0.782 & 7.56 $\pm$ 0.61 $\pm$ 0.52 & 4.13 $\pm$ 0.52 $\pm$ 0.32\\
 $[0.3,0.5]$ & 0.380 & $[0.15,0.2]$ & 0.174 & 0.791 & 3.76 $\pm$ 0.55 $\pm$ 0.39 & 2.04 $\pm$ 0.50 $\pm$ 0.25\\
 $[0.3,0.5]$ & 0.380 & $[0.2,0.3]$ & 0.245 & 0.789 & 4.24 $\pm$ 0.47 $\pm$ 0.39 & 2.44 $\pm$ 0.42 $\pm$ 0.24\\
 $[0.3,0.5]$ & 0.379 & $[0.3,0.5]$ & 0.379 & 0.788 & 5.14 $\pm$ 0.54 $\pm$ 0.41 & 2.84 $\pm$ 0.45 $\pm$ 0.25\\
 $[0.3,0.5]$ & 0.379 & $[0.5,0.9]$ & 0.612 & 0.781 & 7.70 $\pm$ 0.82 $\pm$ 0.56 & 3.46 $\pm$ 0.57 $\pm$ 0.34\\
 $[0.5,0.9]$ & 0.612 & $[0.15,0.2]$ & 0.174 & 0.784 & 4.75 $\pm$ 0.63 $\pm$ 0.55 & 2.39 $\pm$ 0.57 $\pm$ 0.35\\
 $[0.5,0.9]$ & 0.611 & $[0.2,0.3]$ & 0.245 & 0.782 & 8.65 $\pm$ 0.63 $\pm$ 0.52 & 4.57 $\pm$ 0.52 $\pm$ 0.32\\
 $[0.5,0.9]$ & 0.612 & $[0.3,0.5]$ & 0.379 & 0.781 & 9.74 $\pm$ 0.92 $\pm$ 0.57 & 5.01 $\pm$ 0.63 $\pm$ 0.34\\
 $[0.5,0.9]$ & 0.615 & $[0.5,0.9]$ & 0.615 & 0.777 & 12.83 $\pm$ 1.54 $\pm$ 0.81 & 5.06 $\pm$ 0.85 $\pm$ 0.46\\
\toprule
$z_1$ & $\langle z_1\rangle$ & $z_2$ & $\langle z_2\rangle$ &  $ \frac{\langle\sin^2\theta_{2}\rangle}{\langle1+\cos^2\theta_{2}\rangle}$ & \aul\ ($10^{-2}$) & \auc\ ($10^{-2}$) \\ \toprule
$[0.15,0.2]$ & 0.174 & $[0.15,0.2]$ & 0.174 & 0.732 & 0.70 $\pm$ 0.78 $\pm$ 0.36 & 0.37 $\pm$ 0.69 $\pm$ 0.21\\
 $[0.15,0.2]$ & 0.174 & $[0.2,0.3]$ & 0.246 & 0.736 & 0.48 $\pm$ 0.61 $\pm$ 0.32 & 0.23 $\pm$ 0.55 $\pm$ 0.17\\
 $[0.15,0.2]$ & 0.174 & $[0.3,0.5]$ & 0.380 & 0.748 & 1.04 $\pm$ 0.53 $\pm$ 0.31 & 0.52 $\pm$ 0.48 $\pm$ 0.17\\
 $[0.15,0.2]$ & 0.174 & $[0.5,0.9]$ & 0.612 & 0.752 & 3.66 $\pm$ 0.64 $\pm$ 0.39 & 2.00 $\pm$ 0.56 $\pm$ 0.25\\
 $[0.2,0.3]$ & 0.246 & $[0.15,0.2]$ & 0.174 & 0.727 & 1.68 $\pm$ 0.63 $\pm$ 0.32 & 0.93 $\pm$ 0.56 $\pm$ 0.17\\
 $[0.2,0.3]$ & 0.245 & $[0.2,0.3]$ & 0.246 & 0.733 & 1.47 $\pm$ 0.50 $\pm$ 0.29 & 0.79 $\pm$ 0.45 $\pm$ 0.14\\
 $[0.2,0.3]$ & 0.245 & $[0.3,0.5]$ & 0.380 & 0.746 & 1.89 $\pm$ 0.46 $\pm$ 0.31 & 0.98 $\pm$ 0.42 $\pm$ 0.15\\
 $[0.2,0.3]$ & 0.245 & $[0.5,0.9]$ & 0.611 & 0.750 & 2.47 $\pm$ 0.57 $\pm$ 0.35 & 1.26 $\pm$ 0.50 $\pm$ 0.21\\
 $[0.3,0.5]$ & 0.380 & $[0.15,0.2]$ & 0.174 & 0.724 & 1.19 $\pm$ 0.55 $\pm$ 0.31 & 0.67 $\pm$ 0.50 $\pm$ 0.17\\
 $[0.3,0.5]$ & 0.380 & $[0.2,0.3]$ & 0.245 & 0.731 & 1.86 $\pm$ 0.46 $\pm$ 0.31 & 1.01 $\pm$ 0.42 $\pm$ 0.15\\
 $[0.3,0.5]$ & 0.379 & $[0.3,0.5]$ & 0.379 & 0.744 & 2.60 $\pm$ 0.53 $\pm$ 0.30 & 1.39 $\pm$ 0.44 $\pm$ 0.15\\
 $[0.3,0.5]$ & 0.379 & $[0.5,0.9]$ & 0.612 & 0.749 & 2.96 $\pm$ 0.75 $\pm$ 0.37 & 1.30 $\pm$ 0.56 $\pm$ 0.22\\
 $[0.5,0.9]$ & 0.612 & $[0.15,0.2]$ & 0.174 & 0.717 & 1.57 $\pm$ 0.63 $\pm$ 0.39 & 0.80 $\pm$ 0.57 $\pm$ 0.26\\
 $[0.5,0.9]$ & 0.611 & $[0.2,0.3]$ & 0.245 & 0.725 & 2.48 $\pm$ 0.57 $\pm$ 0.35 & 1.18 $\pm$ 0.50 $\pm$ 0.21\\
 $[0.5,0.9]$ & 0.612 & $[0.3,0.5]$ & 0.379 & 0.738 & 4.20 $\pm$ 0.86 $\pm$ 0.37 & 2.01 $\pm$ 0.60 $\pm$ 0.23\\
 $[0.5,0.9]$ & 0.615 & $[0.5,0.9]$ & 0.615 & 0.745 & 5.53 $\pm$ 1.43 $\pm$ 0.52 & 2.05 $\pm$ 0.81 $\pm$ 0.35\\
\toprule 
\end{tabular}
\caption {Light quark (\uds) Collins asymmetries obtained by fitting the U/L and U/C double ratios
s a function of \zbin\ for \kpi hadron pairs in the RF12 frame (upper table) and in the RF0 frame (lower table).
In the first two columns are reported the  $z$ bins and their respective
mean values for the hadron ($K$ or $\pi$) in one hemisphere; in the following two columns, 
the same variables for the second hadron ($K$ or $\pi$) are shown; in the fifth column is summarized 
the ratio of mean values  $\langle\sin^2\theta_{th(2)}\rangle/\langle 1+\cos^2\theta_{th(2)}\rangle$,
and the asymmetry results are reported in the last two columns.
The quoted errors are statistical and systematic, respectively.
The mean values of the quantities reported in the table are calculated by summing the corresponding values for
each \kpi pair and dividing by the number of \kpi pairs that fall into each \zbin\ interval.
Note that the $A^{UL}$ and $A^{UC}$ results are strongly correlated since they are obtained
by using the same data set.}
\label{tab:KPitotZ}
\end{center}
\end{table*}

\begin{table*}[!hbt]
\begin{center}
\begin{tabular}{ c c c c c c c}
\toprule 
\multicolumn{7}{c}{\pipi sample}\\
\toprule 
$z_1$ & $\langle z_1\rangle$ & $z_2$ & $\langle z_2\rangle$ & $ \frac{\langle\sin^2\theta_{th}\rangle}{\langle1+\cos^2\theta_{th}\rangle}$ & \aulTh\ ($10^{-2}$) & \aucTh\ ($10^{-2}$) \\ \toprule
 $[0.15,0.2]$ & 0.174 & $[0.15,0.2]$ & 0.174 & 0.791 & 2.64 $\pm$ 0.37 $\pm$ 0.39 & 1.25 $\pm$ 0.26 $\pm$ 0.20\\
 $[0.15,0.2]$ & 0.174 & $[0.2,0.3]$ & 0.244 & 0.789 & 3.72 $\pm$ 0.29 $\pm$ 0.40 & 1.74 $\pm$ 0.21 $\pm$ 0.19\\
 $[0.15,0.2]$ & 0.174 & $[0.3,0.5]$ & 0.378 & 0.786 & 4.06 $\pm$ 0.24 $\pm$ 0.43 & 1.87 $\pm$ 0.19 $\pm$ 0.21\\
 $[0.15,0.2]$ & 0.174 & $[0.5,0.9]$ & 0.617 & 0.781 & 6.26 $\pm$ 0.34 $\pm$ 0.57 & 2.80 $\pm$ 0.23 $\pm$ 0.29\\
 $[0.2,0.3]$ & 0.244 & $[0.15,0.2]$ & 0.174 & 0.789 & 3.76 $\pm$ 0.29 $\pm$ 0.40 & 1.76 $\pm$ 0.21 $\pm$ 0.19\\
 $[0.2,0.3]$ & 0.244 & $[0.2,0.3]$ & 0.244 & 0.788 & 4.69 $\pm$ 0.21 $\pm$ 0.41 & 2.17 $\pm$ 0.17 $\pm$ 0.20\\
 $[0.2,0.3]$ & 0.244 & $[0.3,0.5]$ & 0.377 & 0.785 & 4.99 $\pm$ 0.21 $\pm$ 0.44 & 2.24 $\pm$ 0.16 $\pm$ 0.21\\
 $[0.2,0.3]$ & 0.244 & $[0.5,0.9]$ & 0.617 & 0.780 & 8.27 $\pm$ 0.36 $\pm$ 0.58 & 3.57 $\pm$ 0.22 $\pm$ 0.28\\
 $[0.3,0.5]$ & 0.378 & $[0.15,0.2]$ & 0.174 & 0.786 & 4.53 $\pm$ 0.25 $\pm$ 0.43 & 2.08 $\pm$ 0.19 $\pm$ 0.21\\
 $[0.3,0.5]$ & 0.377 & $[0.2,0.3]$ & 0.244 & 0.785 & 4.73 $\pm$ 0.21 $\pm$ 0.44 & 2.12 $\pm$ 0.16 $\pm$ 0.21\\
 $[0.3,0.5]$ & 0.377 & $[0.3,0.5]$ & 0.377 & 0.782 & 6.23 $\pm$ 0.33 $\pm$ 0.48 & 2.70 $\pm$ 0.19 $\pm$ 0.23\\
 $[0.3,0.5]$ & 0.378 & $[0.5,0.9]$ & 0.619 & 0.777 & 9.47 $\pm$ 0.59 $\pm$ 0.62 & 3.85 $\pm$ 0.29 $\pm$ 0.30\\
 $[0.5,0.9]$ & 0.617 & $[0.15,0.2]$ & 0.174 & 0.781 & 6.58 $\pm$ 0.37 $\pm$ 0.58 & 2.94 $\pm$ 0.24 $\pm$ 0.29\\
 $[0.5,0.9]$ & 0.617 & $[0.2,0.3]$ & 0.244 & 0.780 & 7.45 $\pm$ 0.35 $\pm$ 0.58 & 3.21 $\pm$ 0.22 $\pm$ 0.28\\
 $[0.5,0.9]$ & 0.619 & $[0.3,0.5]$ & 0.378 & 0.777 & 8.77 $\pm$ 0.59 $\pm$ 0.62 & 3.55 $\pm$ 0.29 $\pm$ 0.30\\
 $[0.5,0.9]$ & 0.622 & $[0.5,0.9]$ & 0.622 & 0.772 & 18.46 $\pm$ 1.31 $\pm$ 0.98 & 6.93 $\pm$ 0.54 $\pm$ 0.43\\
\toprule
$z_1$ & $\langle z_1\rangle$ & $z_2$ & $\langle z_2\rangle$ &  $\frac{\langle\sin^2\theta_{2}\rangle}{\langle1+\cos^2\theta_{2}\rangle}$ & \aul\ ($10^{-2}$) & \auc\ ($10^{-2}$) \\ \toprule
$[0.15,0.2]$ & 0.174 & $[0.15,0.2]$ & 0.174 & 0.725 & 1.26 $\pm$ 0.31 $\pm$ 0.22 & 0.59 $\pm$ 0.24 $\pm$ 0.14\\
 $[0.15,0.2]$ & 0.174 & $[0.2,0.3]$ & 0.244 & 0.735 & 1.66 $\pm$ 0.25 $\pm$ 0.21 & 0.78 $\pm$ 0.20 $\pm$ 0.13\\
 $[0.15,0.2]$ & 0.174 & $[0.3,0.5]$ & 0.378 & 0.744 & 1.41 $\pm$ 0.22 $\pm$ 0.22 & 0.65 $\pm$ 0.18 $\pm$ 0.13\\
 $[0.15,0.2]$ & 0.174 & $[0.5,0.9]$ & 0.617 & 0.750 & 2.39 $\pm$ 0.30 $\pm$ 0.27 & 1.05 $\pm$ 0.22 $\pm$ 0.18\\
 $[0.2,0.3]$ & 0.244 & $[0.15,0.2]$ & 0.174 & 0.722 & 1.52 $\pm$ 0.25 $\pm$ 0.21 & 0.71 $\pm$ 0.20 $\pm$ 0.13\\
 $[0.2,0.3]$ & 0.244 & $[0.2,0.3]$ & 0.244 & 0.733 & 2.12 $\pm$ 0.20 $\pm$ 0.21 & 0.98 $\pm$ 0.16 $\pm$ 0.12\\
 $[0.2,0.3]$ & 0.244 & $[0.3,0.5]$ & 0.377 & 0.742 & 2.13 $\pm$ 0.20 $\pm$ 0.21 & 0.96 $\pm$ 0.16 $\pm$ 0.13\\
 $[0.2,0.3]$ & 0.244 & $[0.5,0.9]$ & 0.617 & 0.748 & 3.03 $\pm$ 0.30 $\pm$ 0.25 & 1.31 $\pm$ 0.20 $\pm$ 0.16\\
 $[0.3,0.5]$ & 0.378 & $[0.15,0.2]$ & 0.174 & 0.718 & 1.58 $\pm$ 0.23 $\pm$ 0.22 & 0.73 $\pm$ 0.18 $\pm$ 0.13\\
 $[0.3,0.5]$ & 0.377 & $[0.2,0.3]$ & 0.244 & 0.729 & 2.09 $\pm$ 0.20 $\pm$ 0.21 & 0.93 $\pm$ 0.16 $\pm$ 0.13\\
 $[0.3,0.5]$ & 0.377 & $[0.3,0.5]$ & 0.377 & 0.738 & 2.64 $\pm$ 0.27 $\pm$ 0.22 & 1.13 $\pm$ 0.17 $\pm$ 0.13\\
 $[0.3,0.5]$ & 0.378 & $[0.5,0.9]$ & 0.619 & 0.746 & 3.89 $\pm$ 0.52 $\pm$ 0.28 & 1.58 $\pm$ 0.27 $\pm$ 0.18\\
 $[0.5,0.9]$ & 0.617 & $[0.15,0.2]$ & 0.174 & 0.712 & 2.20 $\pm$ 0.30 $\pm$ 0.27 & 0.98 $\pm$ 0.22 $\pm$ 0.18\\
 $[0.5,0.9]$ & 0.617 & $[0.2,0.3]$ & 0.244 & 0.724 & 3.06 $\pm$ 0.31 $\pm$ 0.25 & 1.32 $\pm$ 0.21 $\pm$ 0.16\\
 $[0.5,0.9]$ & 0.619 & $[0.3,0.5]$ & 0.378 & 0.735 & 3.76 $\pm$ 0.51 $\pm$ 0.28 & 1.52 $\pm$ 0.27 $\pm$ 0.18\\
 $[0.5,0.9]$ & 0.622 & $[0.5,0.9]$ & 0.622 & 0.743 & 6.76 $\pm$ 1.06 $\pm$ 0.41 & 2.54 $\pm$ 0.46 $\pm$ 0.26\\
\toprule 
\end{tabular}
\caption {Light quark (\uds) Collins asymmetries obtained by fitting the U/L and U/C double ratios
as a function of \zbin\ for kaon pairs in the RF12 frame (upper table) and in the RF0 frame (lower table).
In the first two columns are reported the  $z$ bins and their respective
mean values for the kaon in one hemisphere; in the following two columns, 
the same variables for the second kaon are shown; in the fifth column is summarized 
the ratio of mean values  $\langle\sin^2\theta_{th(2)}\rangle/\langle 1+\cos^2\theta_{th(2)}\rangle$,
and the asymmetry results are reported in the last two columns.
The quoted errors are statistical and systematic, respectively.
The mean values of the quantities reported in the table are calculated by summing the corresponding values for
each \pipi pair and dividing by the number of \pipi pairs that fall into each \zbin\ interval.
Note that the $A^{UL}$ and $A^{UC}$ results are strongly correlated since they are obtained
by using the same data set.} 
\label{tab:PiPitotZ}
\end{center}
\end{table*} 

\begin{figure*}[!ht]
\begin{minipage}[b]{0.65\linewidth}
\centering
\includegraphics[width=\textwidth]{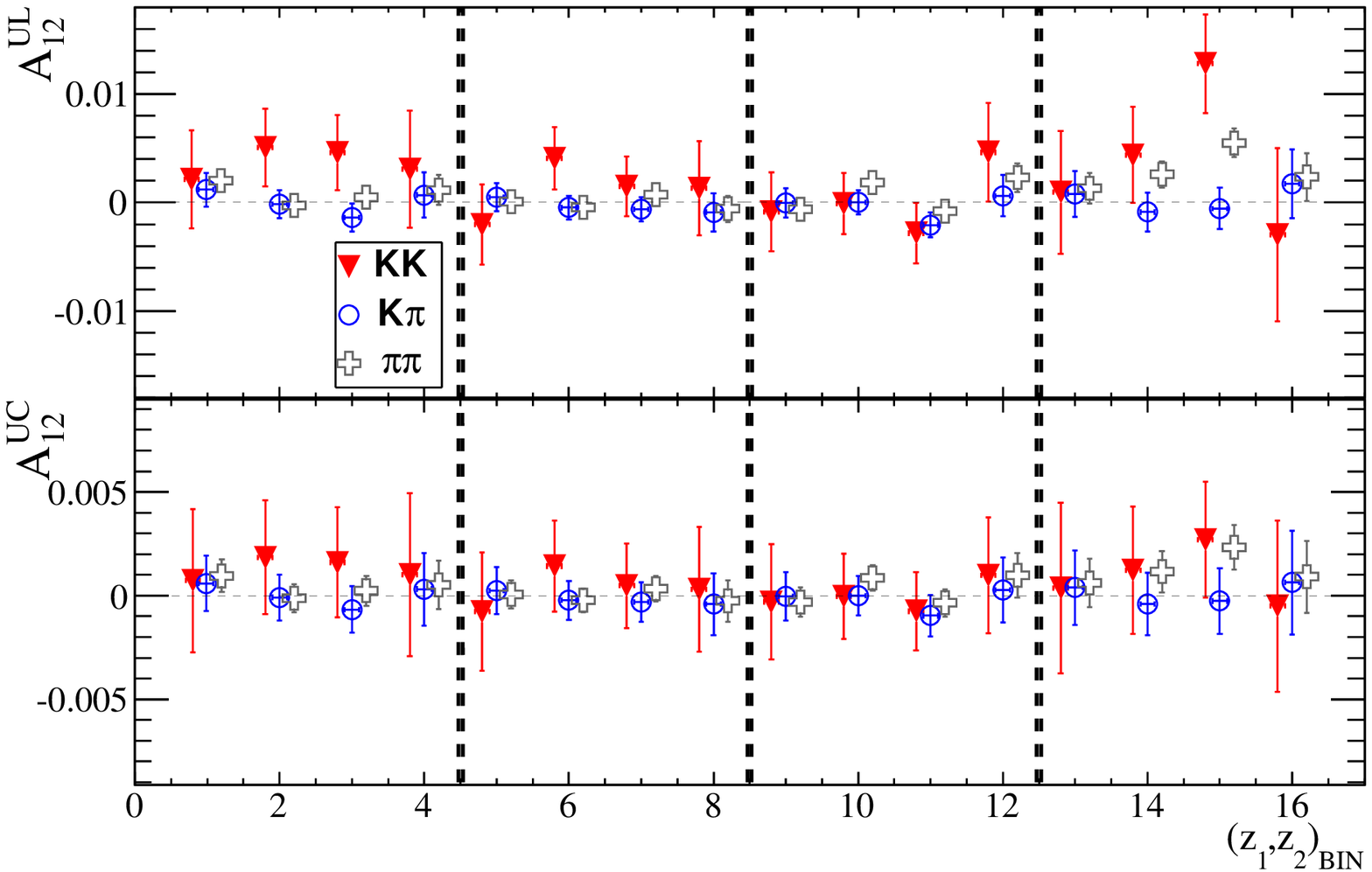}\\
\includegraphics[width=\textwidth]{doublebins}
\end{minipage}\\
\hspace{0.5cm}
\begin{minipage}[b]{0.65\linewidth}
\centering
\includegraphics[width=\textwidth]{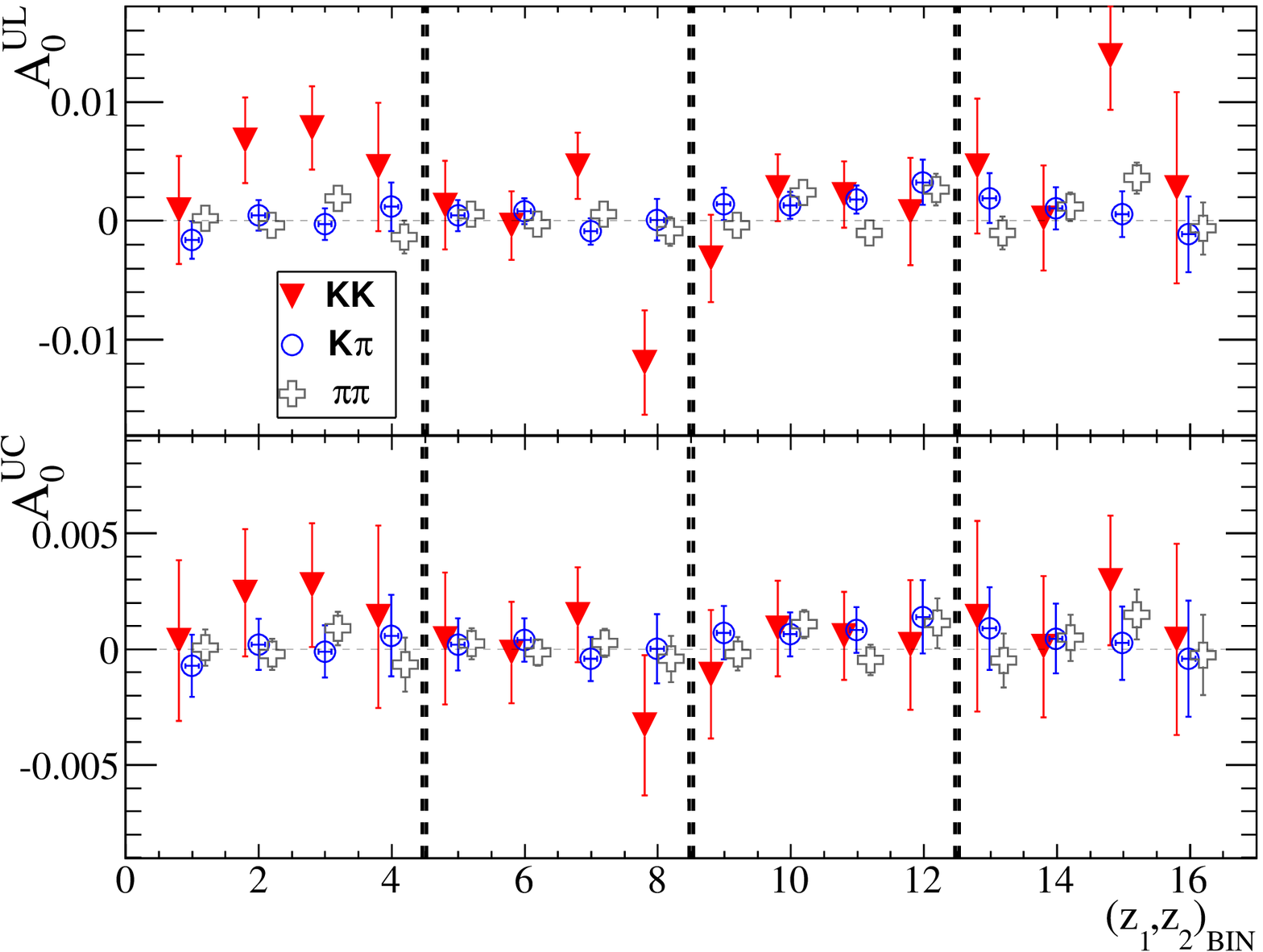}
\includegraphics[width=\textwidth]{doublebins}
\end{minipage}
\caption{ (color online).  Asymmetries measured in the MC sample in RF12 (top)
and RF0 (bottom) for $KK$, $K\pi$, and $\pi\pi$ pairs.
The upper plots show the U/L double ratio, while the lower plots the U/C double ratio.
The 16 ($z_1,z_2$) bins are shown on the x-axis:
in each interval between the dashed lines, $z_1$ is chosen in the
following ranges: $[0.15,0.2]$, $[0.2,0.3]$, $[0.3,0.5]$, and $[0.5,0.9]$, while
within each interval the points correspond to the four bins in $z_2$.
We subtract these biases from the background-corrected asymmetry,
and the statistical errors (represented by the bars around the points) are included into the systematic uncertainties. 
}
\label{fig:bias}
\end{figure*}

\end{document}